# On the Value of the Cosmological Constant in Entropic Gravity

Andreas Schlatter
The Quantum Institute, Gloversville, NY 12078, USA; schlatter.a@bluewin.ch

**Abstract:** We explicitly calculate the value of the cosmological constant, Λ, based on the recently developed theory connecting entropic gravity with quantum-events, induced by transactions, called transactional gravity. We suggest a novel interpretation of the cosmological constant and rigorously show its inverse proportionality to the squared radius of the causal universe $\Lambda \sim R_U^{-2}$.

## 1. Introduction and background

The cosmological constant Λ has played a pivotal role in cosmology ever since Einstein introduced it in 1917. He invented it by the desire to describe a static universe and to counterbalance the effects of gravity. As explained in detail in [1], several phases in the history of Λ can be discerned. Hubble's discovery of the expanding universe lead Einstein to dismiss the cosmological constant in 1931. Already in 1927, however, Lamaître incorporated the comsological constant into his non-static model of the universe interpreting it as a sort of vacuum energy-density, which is still today's standard interpretation [2]. More precise measurements of the Hubble constant in the 1930s again undermined the case for a non-zero Λ. In the 1960s there was a short-lived revival of a non-zero Λ, due to the observation of quasars which seemed to suggest a non-conventional expansion of the universe. Afterwards, physicists thought for a long time that Λ should be exactly zero, but observations by Riess and Perlmutter in 1998 [3,4] on Type 1a supernovae definitively showed that the expansion of the universe is accelerating. This discovery is empirical evidence that $\Lambda > 0$, with a value of $\Lambda \sim 10^{-52} m^{-2}$, as recently calculated by the Planck collaboration [5]. However, there is a colossal mismatch between the theoretical expectations and the empirical facts [6], since calculations of the quantum vacuum-energy lie by hughe orders away from the measured reality [7,8]. This rises the question of the true nature of Λ. A second mystery is the question where the striking relation with the age (size)[1], $R_U$, of the causal universe, namely $\Lambda \sim R_U^{-2}$, comes from [9]. Attempts have been made to explain the origin and value of the constant in e.g. [10 − 15]. Yet, the topic is far from conclusively settled.

Recently, a connection between the relativistic transactional interpretation of quantum mechanics and entropic gravity has been found and a transactional theory of gravity has been developed [16,17]. This theory gives a different physical interpretation of the cosmological constant Λ which, as we will show in this paper, directly leads to the relation $\Lambda \sim R_U^{-2}$. To this end, we first introduce the basics of transactional gravity and then, in section 2, give the main result of this paper, namely the calculation of Λ.

*1.1 Transactional Gravity*

In [16,17] it is shown how empirical spacetime together with its metric structure emerge from quantum-events which happen pairwise by transactions. Quantum systems are elements of a realm of potentialities that empirically get actualized by transactions, which consist of the emission and absorption of (on-shell) photons between these systems. The choice of a propagator thereby defines the direction of time. Empirical spacetime thus becomes the connected set of emission-and absorption events, between which spacetime intervals are being created through the four-momentum of the exchanged photons. Let's look at this in some more detail.

Quantum amplitudes of closed, isolated systems are represented as unit vectors in a Hilbert space, $\psi_x \epsilon H$, also called quantum states. In the transactional interpretation [17] a quantum state $\psi_x$ is launched as an "offer-wave" by an emitter and gets possible responses by "confirmation-waves", represented by dual vectors $\psi_y^*$ launched by possible absorbers. The selection of a specific "response" $\psi_y^*$ is fundamentally indeterministic and leads to a "transaction", which is the actualization of absorption and emission as real events in spacetime, and whose probability (density) is $\left|\delta_{(y-x)} * \psi_x\right|^2 = \left|\psi_y\right|^2$. The relativistic transactional interpretation in addition offers the reason why offer-waves (and confirmation-waves) are actually being

---

[1] Time is expressed by $\tau = ct$ and has hence the dimension of length.

2created, by focusing on the electromagnetic interaction. While emitters correspond to the retarded solutions of the wave equation and to creation operators, absorbers are connected to advanced solutions and to annihilation operators. Relativistic electromagnetic interactions can be thought of as the mutual exchange of virtual photons by quantum fields, creating possibilities in a pre-spacetime process. Transactions, in turn, are characterized by the exchange of real photons and their four-momenta between emitter and absorber. While virtual photons correspond to the Coulomb force and such interactions are unitary, real photons correspond to radiative processes, which are non-unitary interactions ([18], Chapter 5). The general amplitude, $\alpha$, for emission and absorption of real photons is the coupling amplitude between matter-and gauge fields, and the non-unitary transactional process can arise if the conservation laws are satisfied. By this non-unitary exchange of four-momentum the quantum states of emitter and absorber collapse and the physical systems are localized at the corresponding spacetime points (regions).

Empirical spacetime thus becomes the connected set of emission-and absorption points, between which space-time (null)intervals are being created through the four-momenta of the exchanged photons. It is here, where the transactional view touches causal-set theory [19], in which events spread in spacetime by a stochastic Poisson-process. Boson exchange, understood as a decay-process in quantum field theory, is then a special case in this model.[2] Note, that the actualization of a spacetime interval amounts to spontaneously breaking the unitary evolution of the quantum states. At the same time, the exchanged four-momentum selects a space-direction, whereas a time-direction is a priori determined, since only positive energy is being transferred.[3] Because there is no preferred emission-direction, the process is spatially isotropic, and because the whole mechanism is indifferent to specific locations, it is also homogeneous. Observed inhomogeneities of the universe are then the consequence of a possible anisotropic distribution of initial transactions and the spatiotemporal variation of some other parameter, as we will see in section 2.

So far, we have motivated the idea that the formalism of quantum physics is suited to explain the emergence of empirical space and time as unified, yet distinct, dimensions by the mechanism of transactions. Real photon exchange creates metric relations between emitters and absorbers and the mechanism contains therefore an intrinsic way to measure time-intervals by means of the exchanged photons, as described in [17]. This approach clearly lends itself to the relational view of spacetime as it emerges from pairs of emitters and absorbers, and there is no spacetime without matter (although the matter itself is not a component of metrical spacetime; [18], Chapter 8). Mathematically, quantum states can be described as fields parametrized by spacetime coordinates. Note that there is no circularity here, since this description does not suggest that spacetime has a real a priori existence of its own. It is only a model representation of our observations, where we never measure standalone spacetime points. Hence the term "empirical spacetime" in order to distinguish observed reality from mathematical models. We also note that the conception of quantum states as fields over spacetime uses the spacetime parameters as possibilities for localization relative to a particular inertial frame, and that such quantum states are not physically 'in spacetime' where the latter is understood as the emergent manifold of connected events.

By a transaction the involved systems get localized which, in order to keep the entropic balance, leads in a weak limit to Newton's gravitational force. It is mathematically shown in [16,17,20] that this entropic process leads more generally to a gauge of the length in temporal direction, which together with the light-cone structure is sufficient to derive the Einstein equations and hence to govern the four-metric of spacetime [21]. The three-momenta of the exchanged photons in particular lead to a cosmological term in the Einstein equations, which is what we are going to show next.

## 2. The Cosmological Constant

In transactional gravity the energy of the transferred photons gauges the rhythm of becoming in timelike direction as a natural light clock which, together with the light cone structure, leads to Einstein's equation [17,21]. The three-momenta of the transferred photons enter this equation in form of a cosmological constant, $\Lambda$, as we will now show. This fact is intuitively plausible, since the momenta of the photons exercise a repulsive pressure on the material systems involved in the transactions. Note, that in transactional gravity only the energies and momenta of the systems involved in transactions add to the local energy stress tensor and there is no basis for the inclusion of expectation values of quantum fields or alike [16]. Concretely, there arises pressure from a photon three-momentum, emitted in all the spatial directions equiprobably, which defines at a given point the Laue-scalar $T$

$$T = \sum_{i=1}^{3} T_{ii} = \lim_{A_i \to 0} \sum_{i=1}^{3} \frac{F_i}{A_i} = \lim_{A_i \to 0} \sum_{i=1}^{3} \frac{1}{A_i} \frac{dp_i}{dt}. \tag{1}$$

---

[2] The transactional interpretation thinks differently of spacetime than the causal-set approach does, which is unimportant for our purposes.
[3] This amounts to the choice of the Feynmann-propagator as opposed to the Dyson-propagator.

Let $N_R(t)$ be the number of actualizations within (spatial) volume $V_R$ at a time $t$. We have with $x_0 = ct$, $N_R(t) = N_R\left(\frac{x_0}{c}\right) = \widetilde{N}_R(x_0)$ and with the de Broglie relation $|\vec{p}| = \frac{h}{R}$

$$T = -3 \frac{dN_R(t)}{dt} \cdot \frac{1}{A_R} \cdot \frac{h}{R} = -3 \frac{c \cdot h}{3} \cdot \frac{d\widetilde{N}_R(x_0)}{dx_0 V_R} = -c \cdot h \cdot \frac{d\lambda(x_0)}{dx_0}. \tag{2}$$

The negative sign indicates the repulsive effect and the function $\lambda(x_0) = \frac{\widetilde{N}_R(x_0)}{V_R}$ denotes the number of transactional events per spatial volume at time $x_0$. The term $\frac{d\lambda(x_0)}{dx_0}$ is therefore the change-rate of the spatial transactional event-density. In (2) we assumed that $\lambda(x_0)$ is constant over space, which also amounts to the homogoneity and isotropy of space with respect to transactional events. Equation (2) also tacitly assumes that $\lambda(x_0)$ is a differentiable function in $x_0$. This is an assumption, which cannot hold in quantum-mechanics, since quantum-events represent discrete sets and are not deterministic, but obey a random process. The only known Lorentz-invaraint stochastic process for the spreading of quantum events in Minkowski space, such that the number of events is proportionate to the volume, is a Poisson process with constant average (photon) transaction-density rate $\varrho_\gamma$ [22]. Hence, in the above terminology we have for the average event-density $\bar\lambda(x_0)$ and for $\Delta x_0 > 0$

$$\bar\lambda(x_0 + \Delta x_0) = \bar\lambda(x_0) + \varrho_\gamma \cdot \Delta x_0. \tag{3}$$

So, by (3) we can define, in analogy to (2)

$$T_\gamma = -3 \frac{c \cdot h}{3} \cdot \frac{\Delta \bar\lambda(x_0)}{\Delta x_0} = -c \cdot h \cdot \varrho_\gamma. \tag{4}$$

Remember the Einstein equation

$$R_{\mu\nu} = \frac{8\pi G}{c^4}\left(T_{\mu\nu} - \frac{1}{2} T g_{\mu\nu}\right), 0 \leq \mu, \nu \leq 3. \tag{5}$$

Equation (4) leads on the right hand side of Einstein's equation, with $l_P = \sqrt{\frac{G\hbar}{c^3}}$ denoting the Planck lengh, to the term

$$\frac{4\pi G}{c^4} T_\gamma = -\frac{4\pi G h}{c^3} \varrho_\gamma = -8\pi^2 l_P^2 \varrho_\gamma. \tag{6}$$

The right hand side of equation (5) consists of the local energy-momentum distribution, whereas expression (6) is global and independent of any local fields. It hence represents a structural component of equation (5). Consequently, putting it to the left hand side of the Einstein equation leaves us with the cosmological constant [4]

$$\Lambda = 8\pi^2 l_P^2 \varrho_\gamma, \tag{7}$$

---

[4] Note that $[\Lambda] = m^{-2}$.

and Einstein's equation in the form

$$R_{\mu\nu} + \Lambda g_{\mu\nu} = \frac{8\pi G}{c^4}\left(T_{\mu\nu} - \frac{1}{2}T g_{\mu\nu}\right), 0 \leq \mu, \nu \leq 3. \quad [5]$$
(8)

*2.1 Spatial information*

Since transactions localize emitters and absorbers there is an entropy production in the process (see Appendix). The resulting information will, of course, depend upon the physical systems involved. In order to estimate the number of transactional events in the absence of the concrete knowledge of the matter fields, we can define a spatial information content and then calculate the number of bits residing in a region of space $\Omega \subset \mathbb{R}^3$, as done in [17] and shown next.

We assume to work in a local inertial frame throughout the following exposition. Let there be a bounded region $\Omega \subset \mathbb{R}^3$ on a spatial hyperplane and a partition by balls

$$\mathcal{B} = \{B_{\varepsilon_n}(x_n)\}_{x_n \in \Omega, \varepsilon_n > 0}, \cup_{x_n} B_{\varepsilon_n}(x_n) = \Omega.$$
(9)

Relative to the partition $\mathcal{B}$, a position-information can be attributed to a quantum system in terms of square-integrable functions over $\Omega$, $\psi(x) \in L^2(\Omega)$, by

$$I^{\mathcal{B}}(\psi) = -\sum_{x_n \in \Omega} p_{x_n} \ln(p_{x_n}), \qquad p_{x_n} = \int_{B_{\varepsilon_n}(x_n)} |\psi(x)|^2 dx.$$
(10)

By multiplication with the Boltzmann constant, $k_B$, we get

$$S^{\mathcal{B}}(\psi) = I^{\mathcal{B}}(\psi) k_B.$$
(11)

We can ask, whether it is possible to take a different perspective and attribute information not to material systems, but to regions or, idealized, single points $x_0 \in \mathbb{R}^3$. A point, $x_0 \in \Omega$, can empirically be associated with matter or not and hence represents in this sense one bit of information. Given a single physical system, $\psi(x) \in L^2(\Omega)$, we can therefore state that the information of the one bit, $x_0 \in \Omega$, with respect to $\psi(x)$ and the partition $\mathcal{B}$ (9)[6] is

$$I^{\mathcal{B}}_{\psi}(x_0) = -[p_{x_0} \ln(p_{x_0}) + (1 - p_{x_0}) \ln(1 - p_{x_0})].$$
(12)

To find a generic definition, we have to account for all possible partitions $\mathcal{B}$ which amounts to taking into account all probabilities, $0 \leq p_{x_0} \leq 1$. Since we always find some bounded $\Omega \subset \mathbb{R}^3$ with $x_0 \in \Omega$, we can define the information $I(x_0)$, $x_0 \in \mathbb{R}^3$, by

$$I(x_0) = -2\int_0^1 p_{x_0} \ln(p_{x_0}) dp_{x_0} = \frac{1}{2}.$$
(13)

Evidently, (13) is not only independent of a chosen partition $\mathcal{B}$, but also of the particular material system $\psi(x)$. While the choice of a particular $\mathcal{B}$ is, of course, frame-dependent, the described process will lead to the definition of $I(x_0)$ by equation (13) in every local inertial frame.

---

[5] Therefore, the right hand side of Einstein's equation consits in transactional gravity purely of the local energy-momentum of material absorbers.
[6] We pick the ball $B_{\varepsilon_{\tilde{n}}}(x_{\tilde{n}})$ with $x_0 \in B_{\varepsilon_{\tilde{n}}}(x_{\tilde{n}})$ and minimal $|x_0 - x_{\tilde{n}}|$.

## 2.2 Transactional density

In order to investigate the behaviour of $\Lambda$ it is, by equation (7), necessary to understand the transactional event-density $\varrho_\gamma$. To do this, we chose the direct approach and estimate the expected number of transactional events and the volume of the universe since its beginning. We employ a simple model of an expanding, flat universe with a constant Hubble parameter $H_0$ in order to make closed calculations possible. Assuming that the Hubble parameter $H_0$ is constant, allows us to use the Hubble radius $R_{H_0} = \frac{c}{H_0}$ to express the age, $t_U$, of the universe which is $t_U = \frac{1}{H_0}$. We further see that for the expansion factor, $a(t)$, we have with $\frac{\dot{a}(t)}{a(t)} = H_0$ and with $a(t_U) = 1$: $a(t) = e^{H_0(t-t_U)}$. The causal universe at any time $t \leq t_U$ is bounded by the particle horizon, $R_P(t)$, which is:

$$R_P(t) = \int_0^t \frac{c d\tau}{a(\tau)} = \int_0^t \frac{c d\tau}{e^{H_0(\tau - t_U)}} = e R_{H_0}(1 - e^{-H_0 t}). \tag{14}$$

In particular, for today's particle horizon, $R_P(t_U)$, we have $R_P(t_U) = R_{H_0}(e-1)$. In the sequel we set $(e-1) \coloneqq \epsilon$. We further remember, as calculated above (13), that a single point $x_0$ in space represents one bit of information with information content $I(x_0) = \frac{1}{2}$. Assuming the Planck length to be a minimal length in nature, the total number of bits, $n_R$, on the surface of a ball of radius $R$ with surface-area $A_R$ amounts to:

$$n_R = \frac{A_R}{2 l_P^2}. \tag{15}$$

By the holographic principle [23], expression (15) represents the maximum information encoded within the ball $B_R$. Furthermore, the fine-structure constant, $\alpha^2$,[7] is the base-probability for a transaction to happen (at most reduced by a factor stemming from a specific amplitude) and we can reasonably assume that the number of transactional events at any time $x_0$ is (maximally) proportionate to the available spatial information.[8] Hence the (maximum) expected number of transactional events within a ball of radius $R$, $B_R$, is:

$$\bar{N}_R = n_R \cdot \alpha^2 = \frac{A_R \alpha^2}{2 l_P^2}. \tag{16}$$

Since the causal universe has been growing to today's particle horizon $R_P(t_U) = \epsilon R_{H_0}$, we can assume that the total amount of transactional events up to today is coded on the surface of radius $R_P(t_U)$ and it is possible by equations (14) and (16) to set for the total expected number of transactional events in the causal universe up to today, $\bar{N}_{R_P(t_U)}$,

$$\bar{N}_{R_P(t_U)} = \frac{4 \pi R_P^2(t_U) \alpha^2}{2 l_P^2} = \frac{2 \pi \epsilon^2 R_{H_0}^2 \alpha^2}{l_P^2}. \tag{17}$$

To get the transactional event-density we further need to calculate the four-volume of the universe up to today. By equation (14) we get:

$$V_{t_U} = \int_0^{t_U} \frac{4\pi}{3} R_P^3(t) c\, dt = \frac{4\pi e^3}{3} R_{H_0}^3 \int_0^{t_U} (1 - e^{-H_0 t})^3 c\, dt. \tag{18}$$

By the substitution $x = H_0 t$ and by denoting the resulting integral by $I$ [9], equation (18) turns into

$$V_{t_U} = \frac{4\pi e^3}{3} R_{H_0}^3 \frac{c}{H_0} \int_0^1 (1-e^{-x})^3 dx = \left(\frac{4\pi e^3 I}{3}\right) R_{H_0}^4. \tag{19}$$

---

[7] There holds with $q$ denoting the elementary electric charge and $\varepsilon_0$ the dielectrical constant: $\alpha^2 = \frac{q^2}{4\pi \varepsilon_0 \hbar c} \approx \frac{1}{137}$.
[8] This holds because emitting or absorbing material systems are much larger in size than the Planck length.
[9] $I \approx 0.084$.

By setting $C_0 = \left(\frac{4\pi e^3 I}{3}\right)$ which is a constant independent of physical quantities, we have $V_{t_U} = C_0 R_{H_0}^4$. For the average density of expected transactions today, $\varrho_\gamma$, we therefore arrive by equation (16) with a new constant $C_1$ at:

$$\varrho_\gamma = \frac{\overline{N}_{R_P(t_U)}}{V_{t_U}} = \frac{C_1}{l_P^2} \frac{\alpha^2}{R_{H_0}^2}. \tag{20}$$

For $C_1$ we have $C_1 = \left(\frac{3\epsilon^2}{2e^3 I}\right)$. Finally, this leads by the definition of the cosmological constant in equation (7) and by equation (14) with $C_2 = \epsilon^2 C_1 = \frac{3(e-1)^4}{2Ie^3}$

$$\Lambda = 8\pi^2 C_2 \frac{\alpha^2}{R_P^2(t_U)}. \tag{21}$$

In equation (21) we directly recover the measured fact that $\Lambda \sim R_U^{-2}$.[10] Also the key role of the fine-structure constant becomes clear. It is a governing factor of the expected number of transactional events in the universe and as such enters the formula for its expansion.

### 3. Conclusion

The problem of the empirically found tiny value of the cosmological constant has been bothering physicists for a long time. In addition, the proportionality to the squared inverse of the age of the universe seemed a coincidence, albeit an intriguing one. In the theory of transactional gravity, where spacetime and its metric emerge from quantum-events, called transactions, the cosmological constant arises very naturally as the repulsive pressure generated by the three momenta of event-radiation, i.e. of the photons constituting transactions. By the same entropic considerations that lead to an entropic force, i.e. gravity, we also arrive at a natural expression for the cosmological constant which turns out to have exactly the desired behaviour. In our theory $\Lambda$ is at any stage related to the age of the universe but does not necessarily become infinite as time returns back to the origin, since the number of transactions also decreases. In addition, it might be the case that the transactional density-rate regionally differs which would, next to the distribution of initial transactional events, lead to observable inhomogeneous structures at large scales of the universe. It remains to be seen in the future whether transactional gravity is the model which nature actually follows. In any case, the theory very naturally produces a number of explanations for so far rather elusive facts around gravity.

### Appendix

The main ideas in this appendix can be found in [24]. Let there be a bound sate $\mathcal{B}$ in equilibrium with an environment of temperature $T_0$ and a photon $\gamma$ with energy $E_\gamma = h\nu$ before absorption by the bound state. To model the situation as simply as possible and reasonable, we assume that the wave function $\Psi(x, \vec{x}_j)$, $j\epsilon J \subset \mathbb{N}$, $(x, \vec{x}_j)\epsilon(\mathbb{R}, \mathbb{R}^3)$, of the bound state is factorizable as the product of a center of mass component $\psi(x)$ and an orbital component $Y(\vec{x}_j)$, $j\epsilon J \subset \mathbb{N}$, $\Psi(x, \vec{x}_j) = \psi(x) Y(\vec{x}_j)$ [25]. Since the main contribution of the mass $m > 0$ stems from the nucleus, we may assume that the center of mass component $\psi_{p_0}(x)$ "carries" linear kinetic energy, whereas orbital energy components reside in $Y(\vec{x}_j)$. The bound state together with the photon form a closed system $\Sigma_0$ and under the assumption that the bound state "moves freely" with small momentum-uncertainty, the function $\psi_{p_0}(x)$ can be assumed to be a Gaussian[11]. We will use $\psi_{p_0}(x)$ in order to model and analyze the entropic situation before and after absorption.

---

[10] We have today $R_P(t_U) = 46.5 Gly \approx 4{,}2 \cdot 10^{26} m$.

[11] I.e., $|\psi_{p_0}(x)|^2 \sim \mathcal{N}(\mu_x, \sigma_x)$.

Let us define for any wave-function $\psi \epsilon L^2(]-\infty, \infty[, \mathbb{C})$ the information entropy $I_\psi$ to be

$$I_\psi = -\int_{-\infty}^{\infty} |\psi(s)|^2 \ln|\psi(s)|^2 ds. \tag{A1}$$

The total information-entropy $I_{\psi_{p_0}}^{tot}$ of $\psi_{p_0}(x)$ is then defined by

$$I_{\psi_{p_0}}^{tot} = I_{\psi_{p_0}(x)} + I_{\varphi_{p_0}(p)}, \tag{A2}$$

where $\varphi_{p_0}(p) = \hat{\psi}_{p_0}(p)$ is the conjugate state[12]. By a result of Leipnik [26] there holds for any pair of conjugate variables $\psi(x)$ and $\varphi(p)$ and with Planck's constant, $h$,

$$I_\psi^{tot} = I_{\psi(x)} + I_{\varphi(p)} \geq \ln\left(\frac{he}{2}\right), \tag{A3}$$

with equality in case of Gaussian functions, which we may assume to be a good representation of systems in an equilibrium situation, as mentioned above. Since the bound state is supposed to move freely at a definite momentum, $\varphi_{p_0}(p)$ is highly concentrated around a mean value $\mu_p = p_0$ and there is hence a negative entropy contribution $I_{\varphi_{p_0}(p)} < 0$.[13] By (A2) there holds

$$I_{\varphi_{p_0}(p)} = \ln\left(\frac{he}{2}\right) - I_{\psi_{p_0}(x)}. \tag{A4}$$

At the same time, the momentum of the photon $\gamma$ is known to be $p_\gamma = \frac{h\nu}{c}$ and its position is undefinable, since there is no rest-frame. So, we set

$$I_\gamma^{tot} = 0.\tag{A5}$$ [14]

For the total system-entropy $I_{\Sigma_0}^{tot}$ before absorption we therefore have

$$I_{\Sigma_0}^{tot} = I_{\psi_{p_0}}^{tot}. \tag{A6}$$

Let us finally define, in analogy to Boltzmann's *H*-function, the thermodynamic entropy of the system $\Sigma_0$ by

$$S^{\Sigma_0} = k_B I_{\varphi_{p_0}(p)} = -k_B \int_{-\infty}^{\infty} |\varphi_{p_0}(p)|^2 \ln|\varphi_{p_0}(p)|^2 dp, \tag{A7}$$

where $k_B$ denotes the Boltzmann constant. If initially we have $\psi_{p_0}(x) = \psi_{p_0}(x, 0)$ and $\varphi_{p_0}(p) = \varphi_{p_0}(p, 0)$, respectively, then a free evolution leads after some time $t > 0$ to new states $\psi_{p_0}(x, t)$ and $\varphi_{p_0}(p, t)$, still conjugates of each other. [15]The evolution is unitary and causes an increasing dispersion, $\sigma_x(t)$, of the density $|\psi_{p_0}(x, t)|$ around some evolving position-mean value $\mu_x(t)$, while the density $|\varphi_{p_0}(p, t)|$ remains equally concentrated around $p_0$ and $|\varphi_{p_0}(p, t)| = |\varphi_{p_0}(p, 0)|$. Therefore, there holds by definition (A7) for $t \geq 0$

$$S^{\Sigma(t)} = const. \tag{A8}$$

In other words, the entropy of the unitarily evolving free bound state remains constant, as expected from a reversible process.[16]

---

[12] The Fourier-transformed state.
[13] Note that the differential entropy $I_\mathcal{N}$ of a Gaussian $\mathcal{N}(\mu, \sigma)$ is $I_\mathcal{N} = \ln(\sqrt{2\pi}\sigma) + \frac{1}{2}$ and hence $\lim_{\sigma \searrow 0} I_\mathcal{N} = -\infty$.
[14] Consistent with $\mu(p_\gamma) = 1$.
[15] Note that $\psi_{p_0}(x, t)$ is no more a function with real variance.
[16] Time reversal $t \to -t$ demands $\psi \to \psi^*$.

Let us now look at the situation after the absorption. The absorption at some time $t_1 > 0$ does two things at once: it annihilates the photon and localizes the center of mass component and thus transforms system $\Sigma_0$ into a spatially localized system $\Sigma_1$. This leaves us with a state $\psi_{x_1}(x)$, which is a Gaussian well concentrated around some spatial mean value $\mu_x = x_1$. So, there is now a negative entropy contribution $I_{\psi_{x_1}(x)} < 0$ to total entropy (A2). But because of (A3) the entropy-contribution of the conjugate Gaussian $\varphi_{x_1}(p)$ must compensate and we have in analogy to (A4)

$$I_{\varphi_{x_1}(p)} = ln\left(\frac{he}{2}\right) - I_{\psi_{x_1}(x)}. \tag{A9}$$

So, by (A4), (A8) and (A9) the transition $\Sigma(t) \to \Sigma_1$ induces for $0 \leq t < t_1$ an entropy difference of

$$\Delta_{\Sigma(t)}^{\Sigma_1} S = k_B\left(I_{\varphi_{x_1}(p)} - I_{\varphi_{p_0}(p,t)}\right) = k_B\left(I_{\varphi_{x_1}(p)} - I_{\varphi_{p_0}(p,0)}\right) = k_B\left(I_{\psi_{p_0}(x)} - I_{\psi_{x_1}(x)}\right) > 0. \tag{A10}$$

After the measurement, the bound state $\Sigma_1$ will again develop freely $\Sigma_1 \to \Sigma_1(t)$ and by equation (A8) the entropy $S^{\Sigma_1(t)}$ remains constant, while the position state disperses around a moving mean-position.

This research received no funding

1. Rugh, S.E.; Zinkernagel, H. The quantum vacuum and the cosmological constant problem. *Studies in History and Philosophy of Science Part B*: *Studies in History and Philosophy of modern Physics,* 2002, 33(4), 663-705.
2. Lemaître, G. Evolution of the Expanding Universe. *Proc.Nat.Acad.Sci., 20,* 1934, 12.
3. Riess, A.G.; et al. *Astron. J., 116,* 1998, 116.
4. Perlmutter, S.; et al. *Astrophys. J., 517,* 1999, 565.
5. Planck Collaboration, Planck 2015 results, XIII Cosmological Parameters. *Astronomy&Astrophysics,* 2016, 594.
6. Straumann, N. The History of the cosmological constant problem. 18th IAP Colloquium on the Nature of Dark Energy: Observation and Theoretical Results on the Accelerating Universe, 2002.
7. Straumann, N. The mystery of the cosmic vacuum energy densityand the accelerated expansion of the universe. *Astro-ph/9908342.*
8. Weinberg, S. The cosmological constant problem. *Review of Modern Physics,*1989, 61, 1-23.
9. Abbott, L. The mystery of the Cosmological Constant. *Scientific American,* May 1988, 82-88.
10. Zel'dovich, Y.B. The Cosmological Constant and the Theory of Elementary Particles. *Soviet Physics Uspekhi,* 1968, 11, 381-93.
11. Zel'dovich, Y.B. Cosmological Constant and Elementary Particles. *JETP letters,* 1967, 6, 316-17.
12. Barrow, John D.; Douglas, Shaw J. The value of the cosmological constant. *General Relativity and Gravitation, 43,* 2011, 2555-2560.
13. Weinberg, S. Theories of the Cosmological Constant, in Turok, N.G. (edt) *Critical Dialogues in Cosmology,* 1997, (Singapore: World Scientific), 1-10-
14. Sorkin, Rafael D. Is the cosmological "constant" a nonlocal quantum residue of discreteness of the causal set type? *13th international Symposium on Particles, Strings and Cosmology Conference Proceedings,* Vol. 957, 2007, 142-53.
15. Kastner Ruth, E.; Kauffmann, S. Are Dark Energy and Dark Matter different Aspects of the same Physical Process? *Front. in Phys.,* 6, 2018, 71.
16. Schlatter, A. On the Foundations of Space and Time by Quantum-Events. *Found.Phys.,* 52, 7, 2022.
17. Schlatter, A.; Kastner, R. E. (2023) "Gravity from Transactions: Fulfilling the Entropic Gravity Program." *J. Phys. Commun.* 7, 065009.
18. Kastner, R.E. The Transactional Interpretation of Quantum Mechanics: A Relativistic Treatment. *Cambridge University Press*, 2022.
19. Sorkin, R.D. Causal Sets:Discrete Gravity (Notes for the Valdivia Summer School), In Proceedings of the Valdivia Summer School, 2003, ed. A. Goberoff and D. Marlof. https://arxiv.org/abs/gr-qc/0309009.
20. Verlinde, E. On the origin of gravity and the laws of Newton. *JHEP*, 4, 29, 2011.
21. Wald, R.M. General Relativity. *Chicago University Press,* 1984, Appendix D.
22. Bombelli, L.; Henson, J.; Sorkin, R.D. Discreteness without symmetry braking: a theorem. *Modern Physics Letters A,* 24(32), 2009, 2579-2587.
23. Bousso, R. The Holographic principle. 2002, *Rev.Mod.Phys. 74, 825-74.*
24. Kastner, R.; Schlatter, A. Energy Cost of "Erasure" in Physically Irreversible Processes. *Mathematics,* 2014, 12(2), 206.
25. Born, M.; Oppenheimer, R. Zur Quantentheorie der Moleküle. *Annalen der Physik*, 1927, 389 (20), 457–484.
26. Leipnik, R. Entropy and the Uncertainty Principle. *Information and Control*, 1959, 2, 64-79.